\documentclass[a4paper,12pt]{article}
\usepackage[utf8]{inputenc}
\usepackage{cancel}
\usepackage{tikz}
\usepackage{ulem}
\usepackage{amsfonts}
\usepackage{amssymb}
\usepackage{graphicx}
\usepackage{amsmath}
\usepackage{enumerate}
\usepackage{mathtools}
\usepackage{subfig}
\usepackage{color}
\usepackage{tikz}
\usepackage{float}
\usepackage{here}
\usepackage{cite}
\usepackage{mathrsfs}
\usepackage{float,epsfig}
\usepackage{dcolumn}
\usepackage{graphicx}
\usepackage{bm}
\usepackage{amsmath,amssymb,amsthm}
\usepackage[colorlinks=true,linkcolor=blue,citecolor=red]{hyperref}
\usepackage{multirow}
\usepackage{tikz}
\usepackage{tikz-feynman}

\usetikzlibrary{arrows.meta}
\usetikzlibrary{bending}
\usetikzlibrary{calc}
\newcommand{\be}{\begin{equation}}
\newcommand{\ee}{\end{equation}}
\newcommand{\bea}{\setlength\arraycolsep{2pt} \begin{eqnarray}}
\newcommand{\eea}{\end{eqnarray}}

\def\0{{\sst{(0)}}}
\def\1{{\sst{(1)}}}
\def\2{{\sst{(2)}}}
\def\3{{\sst{(3)}}}
\def\4{{\sst{(4)}}}
\def\5{{\sst{(5)}}}
\def\6{{\sst{(6)}}}
\def\7{{\sst{(7)}}}
\def\8{{\sst{(8)}}}
\def\sst#1{{\scriptscriptstyle #1}}

\makeatletter \@addtoreset{equation}{section}

\definecolor{lime}{HTML}{A6CE39}

\begin{document}

\title{{\normalsize \textbf{\Large On Quark Substructures in an Inspired
Unified Model}}}
\author{ {\small Adil Belhaj$^1$, Salah Eddine Ennadifi$^2$\thanks{%
Authors in alphabetical order.} \hspace*{-8pt}} \\
%EndAName
{\small $^1$ D\'{e}partement de Physique, \'Equipe des Sciences de la
mati\`ere et du rayonnement, ESMaR}\\
{\small Facult\'e des Sciences, Universit\'e Mohammed V de Rabat, Rabat,
Morocco} \\
{\small $^2$ LHEP-MS, Facult\'e des Sciences, Universit\'e Mohammed V de
Rabat, Rabat, Morocco } }
\maketitle

\begin{abstract}
Motivated by the growing attention devoted to the Quantum Chromodynamics
sector of the Standard Model, and the recent observations of non-standard
hadronic states, a possible substructure of quarks $q_{f}\equiv
C_{ij}|k_{i}k_{j}\rangle $ in terms of four colorless bound  {\it hyperparticles}  $%
k_{i(=0,\ldots ,3)}$ is investigated in a $SO(10)$-inspired
unified model. This aims to define a subtle structure of matter as well as
an explanation of some particularities of quarks. Precisely, these
hyperparticles $k_{i}$ are bound by a $SU(3)_{h}$ {\it hypercolor} force under
which they are assumed to be charged. Exploiting certain known data, the
masses and the charges of such fundamental particles are discussed along
with the emerged four extra quarks $q_{f}\equiv \delta
_{ii}|k_{i}k_{i}\rangle $ and three  {\it hyperweak} bosons $W_{h}$.\newline
\end{abstract}

\tableofcontents

% \newpage

\newpage

\section{Introduction}

One of the ancient and the big problems in particle physics is to understand
how matter is formed. The most succeeded achievement, in such field
investigations, is the Standard Model of particle physics (SM) dealing with
strong, weak, and electromagnetic interactions \cite{1,2,3}. It describes
with an elegant way almost all the known experimental phenomena in modern
physics. However, the SM framework seems incomplete after the cumulative
advances in High Energy Physics (HEP) experiments and the precision of the
results. There are many issues to review within the SM. Other than the too
many arbitrary parameters which must be chosen to fit the data, there are
six quarks and six leptons which have been observed. These particles fit
into a three generation scheme with three pairs of quarks and leptons having
the same electromagnetic properties but with increasing masses. The charge
quantization is still a mystery. Effectively, quarks have been shown to
carry fractional electrical charges. In fact, it is not clear why the
electromagnetic charges are quantized. However, the hadrons and the leptons
as well as the gauge bosons have discrete charges. Why there are such
generations of fermions ? and why there is such a big difference in the
masses of the light and the heavy fermions? Recently, the motivation and the
observation of non standard hadronic states consisting of more than three
quarks is a concrete motivation for the existence of a deeper level where
the messy hadronic world can not be easily understood in terms of a few
constituent of $1/2$ spin quark flavors \cite{4,5,6,8,9,10}. All such
aesthetic and structural problems within the SM suggest that the beyond
physics is expected to reside with a more deep fundamental level \cite%
{11,12,13,14,15}.

In the majority of such beyond models, all the observed elementary particles
are considered to be composite. In particular, the preon/rishon model
dealing mainly with the possible substructure of quarks and leptons has been
suggested \cite{11,12,13,14,15,16,17,18,19,20}. However, there can be
variants where only either the quarks or the leptons are composite, but not
both. Such a situation can arise simply because only one type of the
fermions has been treated.

This main objective of this letter is to propose a possible compositeness of
the SM particles in a unified model. More precisely, this work provides a
proposal for the substructure of the known quarks $q_{f=u,d}$ in terms of
more fundamental particles named here hyperparticles $k_{i=0,\ldots }$.
Firstly, we start from a $SO(10)$-inspired unified model which consists of
four colorless hyperparticles denoted $k_{i=0,1,2,3}$ carrying charges under
a hypercolor force keeping them in bound states. Then, we derive the charges
of these fundamental particles such that appropriate bound states of them
reproduce the known SM quark ones. Lastly, we investigate the masses and the
stability behaviors of these hyperparticles. Moreover, we discuss the
possibility of the emerged extra quarks as well as the hyperweak
intermediate bosons $W_{h}$ from the proposed extended gauge symmetry of the
model. We end with a conclusion and further challenges.

\section{Compositeness arguments}

\textbf{Throughout the physics history, the reproduction of elementary 
particles based on the Democritean philosophy inherently drove physicists to
view quarks and leptons as possibly made up of some more fundamental
constituents \cite{10,14,16}. However, given the absence of experimental
evidence of the existence of preons, the reasoning from the rishon model is
likely ignored by most readers as unrelated. This has  led to the belief that the
rishon model, supported by a simple and economic interpretation approach,
should not be treated as a subparticle model. In particular, it has been
proposed that rishons cannot be regarded as subparticles. However,  they  should be
viewed as extended entities, i.e. strictly one-dimensional objects \cite%
{232,233}. In this view, the Rishon model and similar theories can not be
qualified as subparticle models in the traditional meaning of the word. This
could conduct to the possibility of the modification of the concept of
compositeness as well. }Here, despite the disfavors to the idea of ordinary
compositeness mentioned above, we prefer to follow, heuristically, the route
of composite models throughout the physics history based on the Democritean
philosophy by restricting the present compositeness approach to the quarks
only. Indeed, it is known that the large part of the observable matter is
made of strongly interacting quarks and gluons. However, there have been
still arguments that there is something more fundamental behind. Before
going ahead, we quote, among others, the strong quark substructure arguments:

\begin{enumerate}
\item there are six quarks with a specific pattern, i.e., the family
structure;

\item the strong force between quarks is mediated by eight gluons carrying
color-anticolor charges with a specific behavior;

\item a disagreed hint is associated with quark instabilities;

\item the recent observations of non-standard hadronic states consisting of
more than three quarks.

In base of the above arguments, and not only for a sake of simplicity, there
is no particular reason why quarks would be as fundamental as leptons in the
SM and therefore it should be speculated on their compositeness in order to
unveil certain properties associated with the above quark features.
\end{enumerate}

\section{Hyperparticle modeling}

\subsection{Hyperparticles and quark states}

\textbf{Inspired by the six quark flavors and the three corresponding
families, we would like to investigate quark composite models. A priori,
there could be many scenarios reproducing the known quark features. A close
inspection reveals that we could propose at least four elementary particles
with different quantum numbers. Essentially, four hyperparticles }$k_{0}$%
\textbf{, }$k_{1}$\textbf{, }$k_{2}$\textbf{\ and }$k_{3}$\textbf{\ are
needed for a hyperparticle model. Physical considerations in matters of spin
require that such needed fundamental states could correspond to higher-spin
particles introduced in many gravitational supersymmetric extensions, and
envisaged in string-theoretical frameworks \cite{235,236,237}. This could
motivate the stringy origin of the underlying theory of the proposed
subquark model. Having this said, we add that rational and logical reasons
impose extra requirements of possible combinations associated with singlet,
duplet, triplet and quadruplet configurations. It is obvious that the
singlet and the quadruplet arrangements are not relevant. Supported by  such arguments, we could consider the following possible di-hyperparticle
combinations }$\left( k_{i}k_{j}\right) _{i,j}$\textbf{\ such as }%
\begin{eqnarray}
\left( k_{i}k_{j}\right) _{i,j=0,1,2,3}
&=&(k_{0}k_{0}),(k_{0}k_{1}),(k_{0}k_{2}),(k_{0}k_{3})  \notag \\
&&(k_{1}k_{1}),(k_{1}k_{2}),(k_{1}k_{3})  \notag \\
&&(k_{2}k_{2}),(k_{2}k_{3})  \label{eq1} \\
&&(k_{3}k_{3}),  \notag
\end{eqnarray}%
where $k_{i}=0,1,2,3$ are interchangeable. At this level, in the
construction of the model, we need to determine the corresponding quarks of
the di-hyperparticle bound states. For that, we consider the hyperquark
state configurations 
\begin{equation}
q_{f}=C_{ij}|k_{i}k_{j}\rangle ,\qquad i,j=0,1,2,3  \label{eq2}
\end{equation}%
where $f$ is the quark flavor index. $C_{ij}$ is an order 2 tensor, being
exploited to provide a matrix configuration, given by 
\begin{equation}
C_{ii}=\delta _{ii},\qquad C_{i>j}=\epsilon _{ij}  \label{eq3}
\end{equation}%
where $\delta _{ij}$ and $\epsilon _{ij}$ are the diagonal and the
antisymmetric known tensors. In this view, we would expect to have more than
the observed SM six quarks. Actually, this will be revealed later on.

\subsection{Underlying symmetries and charges}

\textbf{At this level, and in a similar spirit to confinement conception of
the model of rishons where the latters are supposed to be bounded into
fundamental fermions by }$SU(3)$\textbf{\ hypercolor inter-preon forces \cite%
{234}, there ought to be certain kind of interactions binding the
hyperparticles from which the color degrees of freedom can emerge. The only
distinction here is that new }$SU(3)$\textbf{\ scheme is supposed to concern
only the quark sector. Thus, analogous to the quark binding force, we call
the hyperparticle binding force the hypercolor force. It is proposed to be a
gauge symmetry with the subindex }$h$\textbf{\ referring to the
corresponding hypercolor charge, under which quarks are neutral, with
{\it hypergluons} }$G_{h}$\textbf{\ as hypercolor force vectors binding the
hyperparticle via hypergluons exchange. Considering the fact that the
hypercolor interaction is thus assumed to be a confining force, the free
hyperparticles would be unobservable even though they might have small
masses. In fact, this symmetry should involve a }$SU(3)$\textbf{\ subgroup
recovering the strong interaction features. Inspired by HEP investigations,
it could be many local gauge symmetries. One scenario may concern a }$SO(10)$%
\textbf{\ symmetry. Indeed, the above particle configurations (\ref{eq2})
and (\ref{eq3}) could hold up the following symmetry factorization} 
\begin{equation}
SO(10)_{h}=SO(4)_{h}\times SO(6)_{h}
\end{equation}%
being equivalent to 
\begin{equation}
SO(10)_{h}=SU(2)_{h}^{2}\times SU(4)_{h}.
\end{equation}%
In this factorization, the first factor indicate that one has two kinds of
doublets. To reproduce the correct color assignment of quarks, all
hyperquarks should be $SU(3)_{h}$ color triplets which could be supported by
the following split symmetries 
\begin{equation}
SU(2)_{h}^{2}\times SU(4)_{h}=SU(2)_{h}^{2}\times U(1)_{h}\times SU(3)_{h}
\label{sym}
\end{equation}%
where the consequences of the additional symmetry $SU(2)_{h}$ will have to
do with the corresponding three extra hypergauge fields (hyperweak gauge
bosons) in the model. This will be exhibited later on. In this view, we can
admit the breaking route down to the SM symmetry and then to the
electromagnetic symmetry in the following scenario

\begin{equation}
SU(2)_{h}^{2}\times U(1)_{h}\times SU(3)_{h}\overset{hyperbreaking}{%
\rightarrow }SU(3)_{C}\times SU(2)_{L}\times U(1)_{Q}\overset{EW\text{ }%
breaking}{\rightarrow }SU(3)_{C}\times U(1)_{Q_{em}}
\end{equation}%
where the hyperbreaking mechanism is unknown.

Regarding quarks as bound states consisting of di-hyperparticle states, and
considering the known six quark charges, a possible set of the hypercolor $%
Q_{h}$ and the electromagnetic $Q_{em}$ charges of the hyperparticles can be
derived. These are listed in table 1.

\begin{center}
\begin{tabular}{|l|l|l|l|l|}
\hline
$k_{i}$ & $k_{0}$ & $k_{1}$ & $k_{2}$ & $k_{3}$ \\ \hline
$Q_{em}$ & $-2/3$ & $1/3$ & $1/3$ & $1/3$ \\ \hline
$Q_{S}$ & \ $3$ & $\ 3$ & $\ 3$ & $\ 3$ \\ \hline
\end{tabular}

\bigskip Tab.1: Hypercolor and electromagnetic charges of the four
hyperparticles.
\end{center}

Thus, the hyperparticle $k_{i}$\ could be distinguished by their charge
numbers. Their masses $m_{k_{i}}$ will be discussed later on. Given these
charges, we identify the quark flavors, the up and the down quarks in terms
of the di-hyperquark combinations as these states 
\begin{eqnarray}
u_{f} &=&C_{ij}|k_{i}k_{j}\rangle _{ij=1,2,3},\text{ \ }  \label{eq6} \\
\text{\ \ }d_{f} &=&\epsilon _{ij}|k_{i}k_{j}\rangle _{i=0,j=1,2,3}.
\label{eq7}
\end{eqnarray}%
These quark identifications could then be illustrated in table 2.

\begin{center}
\begin{tabular}{|l|l|l|l|l|}
\hline
$k_{i}$ & $k_{0}$ & $k_{1}$ & $k_{2}$ & $k_{3}$ \\ \hline
$k_{0}$ & $\left\vert k_{0}k_{0}\right\rangle \equiv x$ & $\left\vert
k_{1}k_{0}\right\rangle \equiv d$ & $\left\vert k_{2}k_{0}\right\rangle
\equiv s$ & $\left\vert k_{3}k_{0}\right\rangle \equiv b$ \\ \hline
$k_{1}$ & $\left\vert k_{0}k_{1}\right\rangle \equiv d$ & $\left\vert
k_{1}k_{1}\right\rangle \equiv y$ & $\left\vert k_{2}k_{1}\right\rangle
\equiv u$ & $\left\vert k_{3}k_{1}\right\rangle \equiv c$ \\ \hline
$k_{2}$ & $\left\vert k_{0}k_{2}\right\rangle \equiv s$ & $\left\vert
k_{1}k_{2}\right\rangle \equiv u$ & $\left\vert k_{2}k_{2}\right\rangle
\equiv z$ & $\left\vert k_{3}k_{2}\right\rangle \equiv t$ \\ \hline
$k_{3}$ & $\left\vert k_{0}k_{3}\right\rangle \equiv b$ & $\left\vert
k_{1}k_{3}\right\rangle \equiv c$ & $\left\vert k_{2}k_{3}\right\rangle
\equiv t$ & $\left\vert k_{3}k_{3}\right\rangle \equiv w$ \\ \hline
\end{tabular}

Tab.2: Ten possible di-hyperparticle states giving rise to six quarks and
four extra ones.
\end{center}

From the table 2, we can read through the other four possible hyperparticle
bound states being four exotic states 
\begin{equation}
q_{f}=C_{ii}|k_{i}k_{i}\rangle =\delta _{ii}|k_{i}k_{i}\rangle \equiv x,y,z,w%
\text{.}
\end{equation}%
To unveil more data, these extra quarks will be investigated in the
fourth-coming paragraphs.

\subsection{Dynamics and mass spectrum}

Considering the fact that quarks are really made up of hyperparticles as
constituents, their fundamental status must therefore be due to some more
intricate dynamics. At the hyperparticle scale $\Lambda _{h}$, the hypercolor 
interaction should then appear among them. Moreover, the dynamics of the
hyperparticles and the hypergluons could be derived from the fundamental
hyperparticle Lagrangian with the underlying sub-symmetry $SU(3)_{h}$. At
energies near or above the quark compositeness scale $\Lambda _{h}$, the
fundamental hyperparticle Lagrangian should read, roughly, as 
\begin{equation}
\mathcal{L}_{k}=\overline{k}_{\alpha }\left[ i\gamma ^{\mu }\left( \partial
_{\mu }\delta _{\alpha \beta }-ig_{S}\left( T_{a}\right) _{\alpha \beta
}\right) \left( G_{\mu }^{a}\right) _{h}-m_{k}\delta _{\alpha \beta }\right]
k_{\beta }-\mathcal{L}_{gauge}  \label{dyn}
\end{equation}%
where now $k_{\alpha }=k_{\alpha }\left( {x^{\mu }}\right) $ is the
dynamical spacetime hyperparticle field in the fundamental representation of
the $SU(3)_{h}$ gauge group indexed by $\alpha ,$ $\beta $ $=1,..3$. $\gamma
^{{\mu }}$ are the Dirac matrices, $g_{h}$ is the hypercoupling strength
coupling superquarks to the hypergluon fields $\left( G_{\mu }^{a}\right)
_{h}$. $T_{a}$ $(a=1,\ldots ,8)$ are the infinitesimal $SU(3)_{h}$
generators in the adjoint representation. The last term $\mathcal{L}_{gauge}$
in eq.(\ref{dyn}) denotes the kinetic energy of the corresponding involved
gauge fields. At energies below the compositeness scale $\Lambda
_{h}\rightarrow \Lambda _{QCD}$, we restrict to the mass terms. Grossly, the
corresponding effective mass Lagrangian is

\begin{equation}
\mathcal{L}_{q}(q_{f})=\sum_{\alpha \beta }\left( m_{q}\right) _{\alpha
\beta }\overline{q}_{f}q_{f^{\prime }}=\sum_{\alpha \beta }\Lambda
_{h}^{-3}\left( m_{kk}\right) _{\alpha \beta }\overline{C_{ij}}|\overline{%
k_{i}}\overline{k_{j}}\rangle C_{ij}|k_{i}k_{j}\rangle .  \label{eq12}
\end{equation}%
In this way, the effective Lagrangian of the di-hyperparticle $%
q_{f}=C_{ij}\left\vert k_{i}k_{j}\right\rangle $ is suppressed by inverse
power of $\Lambda _{h}$. The states $|\overline{k_{i}}\overline{k_{j}}%
\rangle $ and $|k_{i}k_{j}\rangle $ are the Dirac spinors and their adjoints
in terms of the di-hyperparticle states. At this level, although the
involved quark compositeness scale $\Lambda _{h}$ and the mechanism of the
mass generation of the hyperparticles is unknown, their masses could be
investigated from the corresponding quark features. Concretely, from (\ref%
{eq12}), they can be bounded as follows 
\begin{equation}
m_{k_{i}}\ll \frac{m_{|k_{i}k_{j}\rangle }}{2}=\frac{m_{q_{f}}}{2}
\label{eq13}
\end{equation}%
which corresponds to 
\begin{eqnarray}
m_{k_{3}} &\ll &\frac{m_{\left\vert k_{3}k_{2}\right\rangle }}{2}=\frac{m_{t}%
}{2}\text{ \ }  \notag \\
m_{k_{2}} &\ll &\frac{m_{\left\vert k_{2}k_{0}\right\rangle }}{2}=\frac{m_{s}%
}{2}  \notag \\
m_{k_{1}} &\ll &\frac{m_{\left\vert k_{2}k_{1}\right\rangle }}{2}=\frac{m_{u}%
}{2}\text{\ }  \label{eq14} \\
m_{k_{0}} &\ll &\frac{m_{\left\vert k_{1}k_{0}\right\rangle }}{2}=\frac{m_{d}%
}{2}.  \notag
\end{eqnarray}%
These bounds provide the mass ranges in table 3 as

\begin{center}
\begin{tabular}{|l|l|l|l|l|}
\hline
$k_{i}$ & $k_{0}$ & $k_{1}$ & $k_{2}$ & $k_{3}$ \\ \hline
$m_{k_{i}}$ & $m_{k_{0}}\ll 2.4$ $MeV$ & $\ll 1.1$ $MeV$ & $\ll 48$ $MeV$ & $%
\ll 86$ $GeV$ \\ \hline
\end{tabular}%
\\[0pt]
Tab.3: Upper mass bounds of the four hyperparticles.
\end{center}

It follows from the mass bounds eq.(\ref{eq13}) that the hyperparticles might
have small masses. However, they would be unobservable due to the confining
hypercolor interaction.

\subsection{Emerged extra quark states}

From the mass bounds listed in table 3, we turn to the masses of the four
exotic states $x$, $y$, $z$ and $w$. Indeed, they could be approached via
the following inequalities 
\begin{eqnarray}
m_{b_{i}=\left\vert k_{i}k_{i}\right\rangle } &\geqslant
&m_{q_{i}=\left\vert k_{i-1}k_{i}\right\rangle },\text{ }i=3  \notag \\
m_{b_{i}=\left\vert k_{i}k_{i}\right\rangle } &\leqslant
&m_{q_{i}=\left\vert k_{i+1}k_{i}\right\rangle },\text{ }i=1,2  \label{eq15}
\\
m_{b_{i}=\left\vert k_{i}k_{i}\right\rangle } &\leqslant
&m_{q_{i}=\left\vert k_{i+2}k_{1}\right\rangle },\text{ }i=0,  \notag
\end{eqnarray}%
which correspond to 
\begin{eqnarray}
m_{w=\left\vert k_{3}k_{3}\right\rangle } &\geqslant &m_{t=\left\vert
k_{3}k_{2}\right\rangle }\text{ }  \notag \\
m_{z=\left\vert k_{2}k_{2}\right\rangle } &\leqslant &m_{t=\left\vert
k_{3}k_{2}\right\rangle }  \notag \\
m_{y=\left\vert k_{1}k_{1}\right\rangle } &\leqslant &m_{u=\left\vert
k_{2}k_{1}\right\rangle }  \label{eq16} \\
\text{ }m_{x=\left\vert k_{0}k_{0}\right\rangle } &\leqslant
&m_{s=\left\vert k_{2}k_{0}\right\rangle }.  \notag
\end{eqnarray}%
Using the table 1 and the above equations, we obtain the masses and the
charges of such extra emerged states

\begin{center}
\begin{tabular}{|l|l|l|l|l|}
\hline
$\left\vert k_{i}k_{i}\right\rangle $ & $x=\left\vert
k_{0}k_{0}\right\rangle $ & $y=\left\vert k_{1}k_{1}\right\rangle $ & $%
z=\left\vert k_{2}k_{2}\right\rangle $ & $w=\left\vert
k_{3}k_{3}\right\rangle $ \\ \hline
$Q_{b_{I}}$ & $-4/3$ & $2/3$ & $2/3$ & $2/3$ \\ \hline
$m_{b_{I}}$ & $\leqslant 1,3$ $GeV$ & $\leqslant 3$ $MeV$ & $\leqslant 173$ $%
GeV$ & $\geqslant 173$ $GeV$ \\ \hline
\end{tabular}

Tab.4: Mass bounds of the emerged four extra quark states.
\end{center}

According to the mass ranges of these extra quark states, except the fourth
one $w$ which could be more heavier than the weak SM bosons, the remaining
three states $x$, $y$, and $z$ seem to be lighter. It has been believed that
these extra states with such fractional electromagnetic charges and
accessible mass scale are expected to open up for some thrilling
speculations. In particular, this concerns the question of a possible
substructure of the weak bosons $W^{\pm }$ and $Z^{0}$, especially after the
recent observed mass excess-like in the $W$ boson mass. Despite the fact
that there have been many searches for new resonances ($e^{+}e^{-}$ and $pp$
collisions at high energies) and most of them seem to focus on a heavy
fourth family \cite{22,23,230}, the idea that such light extra particles
have escaped discovery is more likely. Indeed, the recent observation of
higher quark combinations, tetraquarks and pentaquarks, at CERN \cite{6,
24,25}, is a strong indication for the existence of more exotic hadrons ($n$%
-quark states). These quark combinations and mass scales could find an
explanation in terms of the exotic quarks listed in table 4.

\section{Hyperparticle stability behavior}

Although we have given the impression throughout this quark compositeness
approach that the hyperparticles are likely stable, like the situation for
most compositeness models of the SM particles \cite{13,14,15,16}, there is,
in fact, no physical reason preventing them from being instable\footnote{%
A quark substructure model must therefore either have absolutely stable
hyperparticle, or rely on yet unstable and composite hyperquarks.}. In such a
case, all weak quark decays in nature are therefore consequences of a deep
mutation of hyperparticles via the three hyperweak gauge bosons $W_{h}^{i=1,2,3}
$ belonging to the hyperweak interactions $SU(2)_{h}$ of the underlying
gauge symmetry (eq.(\ref{sym})) into other ones with lower total rest
masses. For instance, the beta decay $d\rightarrow u+W\rightarrow u+e^{-}+%
\overline{v}_{e}$ reads now at the hyperparticle level as 
\begin{equation}
(k_{1}k_{0})\rightarrow (k_{2}k_{1})+W_{h}\rightarrow
(k_{2}k_{1})+W+...\rightarrow (k_{2}k_{1})+e^{-}+\overline{v}_{e}+...
\label{eq10}
\end{equation}%
where the hyperweak gauge boson $W_{h}$ mediates the mutation of the
hyperparticle decays to the SM weak gauge bosons $W$ with possible unknown
particles. Afterwards, the latters decay to the electron and the
antineutrino. More precisely, we could expect the following decay scenario 
\begin{equation}
k_{0}\rightarrow k_{2}+W_{h}\rightarrow k_{2}+W+...\rightarrow k_{2}+e^{-}+%
\overline{v}_{e}+...  \label{eq11}
\end{equation}%
This is shown in the figure 1.

\begin{figure}[th]
\begin{center}
\begin{tikzpicture}
\begin{feynman}
\vertex (a1) {\(k_0\)};
\vertex[right=3cm of a1] (a2);

\vertex[right=3cm of a2] (a3) {\(k_2\)};

\vertex[right=1.5cm of a3] (a4);

%% See section 13.5 of PGF/TikZ manual
\vertex at ($(a2)!0cm!45:(a2)$) (d);
%% Equivalent way to obtain (d):
% \vertex at ($(b2)!0.5!(b3) + (0, -0.5cm)$) (d);
\vertex[above=of a3] (c1) ;
\vertex[above=2em of c1] (c3) ;
\vertex at ($(c1)!0.5!(c3) - (2.3cm, 0)$) (c2);
\vertex[above=of a4] (n1) {\(e^-\)};
\vertex[above=2em of n1] (n3) {\(\bar{\nu}_e\)} ;
\vertex at ($(n1)!0.62!(n3) - (1.5cm, 0)$) (n2);

\diagram* {
  (a3) --  (a2) -- (a1),

(c3) -- [ boson, edge label=\(W\), out=180, in=45] (c2) -- [ scalar, out=-45, in=180] (c1),
(a2) -- [boson, left] (d),
(d) -- [boson, bend left, edge label=\(W_h\)] (c2),
(n3) -- [ out=180, in=45] (n2) -- [ out=-45, in=180] (n1),
};

\end{feynman}
\end{tikzpicture}
\end{center}
\caption{Beta decay in the subquark model where the dashed line represents
possible unknown particles.}
\label{F1}
\end{figure}
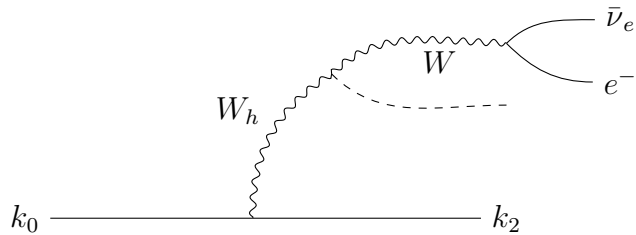

As we can see, accepting the possibility that these hyperparticles are not
assumed to be absolutely stable gives a more fundamental explanation for the
possible decays of quarks. However, the price to pay for this is to find
oneself again faced with the question of the hyperparticle compositeness.
This seems like an endless path toward the ultimate bricks of matter.

\section{Further challenges}

In this work, we have proposed a description of the possible subquark
constituents. It deals with the compositeness of the SM quarks in terms of
hyperparticles in a way that incorporates, in addition to the SM quantum
numbers, a new charge named hypercharge of the $SU(3)_{h}$ symmetry of a
hypercolor interaction which binds hyperparticles together inside quarks as
for the standard quark color charges. In this model, four hyperparticles $%
k_{i}$ ($i=0,1,2,3$), with possibly higher spins, are required as
fundamental subquark particles. By investigating the ten possible
di-hyperparticle combinations, six of them correspond to the SM quarks while
the remaining four ones correspond to four exotic states named $x$, $y$, $z$
and $w$. From the known SM properties, such four hyperparticles have been
identified and likewise those of the four exotic ones. This opens up for
widespread speculations. For the mass spectrum, although the mechanism for
the hyperparticles mass generation is unknown, upper bounds of the
hyperparticle masses as well as of the exotic quarks have been derived from
the corresponding effective di-hyperparticle quarks Lagrangian inspired from
the QCD Lagrangian of quarks. More precisely, from the mass spectrum of the
four hyperparticles, $m_{k_{1}}\leqslant m_{k_{0}}\sim m_{k_{2}}\leqslant
m_{k_{3}}\ll m_{t}/2$ $\sim 86$ $GeV$, the masses of the four extra quarks
have been bounded such as $m_{y}\leqslant m_{x}\leqslant m_{z}\leqslant
m_{t}\leqslant m_{w}$. Except the heavy $w$ hyperparticle mass $m_{w}$, this
mass range (covered in the already accessible energy) should motivate the
appeal of more finer detection techniques.

\textbf{Although this model serves for the understanding of some facts that
are not normally justified within the SM, many problems remain to be solved.
In fact, the model lacks a dynamical basis. This means that the
hyperparticle mass acquisition mechanism is not explained and the
hypercolor force that keeps them together is not well described. It is not
yet clear, at least for us, if all phenomenologically successful aspects of
the electroweak theory can be explained by hyperparticles. Moreover, a
disagreed hint is that the hyperparticles are unstable! Then, where will
this sequence of explanations end? We think, not until we find a stable
constituent level, a lot of work remains to be maintained for a deeper
understanding of the quantitative success of the SM. In this view, where a
stringy nature of quark confinement seems to be potential, we believe that
most of the proposed subparticle models, including ours, could be best
viewed as incomplete images of a certain ultimate theory.}

\section*{Acknowledgements}

The authors would like to thank theirs families for support and patience.
Precisely, they would like to thank their mothers: Fatima (AB) and Fettouma
(SEE).

\end{document}